\documentclass[conference]{IEEEtran}
\usepackage{amsmath}
\usepackage[pdftex]{graphicx}
\usepackage{graphicx}
\usepackage{pstricks}
\usepackage{amsfonts}
\usepackage{tabularx}

\newtheorem{Lemma}{Lemma}
\newtheorem{Remark}{Remark}
\newtheorem{Proposition}{Proposition}

\usepackage{amssymb}
\usepackage{multirow}
\usepackage{epsfig}
\usepackage{epstopdf}

\usepackage[noadjust]{cite} 
\usepackage[noend]{algpseudocode}
\usepackage{algorithm,algpseudocode,graphicx}

\begin{document}
\title{User-Pair Selection for QoS-Aware Secrecy Rate Maximization in  Untrusted NOMA}
\author{\IEEEauthorblockN{Sapna Thapar$^{1}$, Deepak Mishra$^{2}$, Ravikant Saini$^{1}$, and Zhiguo Ding$^{3}$}
\IEEEauthorblockA{$^{1}$Department of Electrical Engineering, Indian Institute of Technology Jammu, Jammu $\&$ Kashmir, India\\
$^{2}$School of Electrical Engineering and Telecommunications, University of New South Wales, Sydney, Australia\\
$^{3}$School of Electrical and Electronic Engineering, University of Manchester, Manchester, United Kingdom\\
Emails: thaparsapna25@gmail.com, d.mishra@unsw.edu.au, ravikant.saini@iitjammu.ac.in, zhiguo.ding@manchester.ac.uk
}}
\maketitle

\begin{abstract}
Non-orthogonal multiple access (NOMA) has been
recognized as one of the key enabling technologies for future generation wireless networks. Sharing the same time-frequency
resource among users imposes secrecy challenges in NOMA in
the presence of untrusted users. This paper characterizes the
impact of user-pair selection on the secrecy performance of
an untrusted NOMA system. In this regard, an optimization
problem is formulated to maximize the secrecy rate of the strong
user while satisfying the quality of service (QoS) demands of
the user with poorer channel conditions. To solve this problem,
we first obtain optimal power allocation in a two-user NOMA
system, and then investigate the user-pair selection problem in
a more generalized four user NOMA system. Extensive performance
evaluations are conducted to validate the accuracy of the
proposed results and present valuable insights on the impact of
various system parameters on the secrecy performance of the
NOMA communication system.
\end{abstract}
\bstctlcite{IEEEexample:BSTcontrol}
\section{Introduction}\label{sec_introduction}
Owning to the exponential growth in the Internet of Things, achieving spectral-efficient communication is an inevitable challenge in wireless networks.  Further, cyber-threats have drastically expanded due to a large number of devices getting connected to the Internet. To address these critical issues, the coexistence of non-orthogonal multiple access (NOMA) and physical-layer security (PLS) is being recognized as a promising solution \cite{ding2016impact}, \cite{8509094}. However, due to significant co-channel interference in NOMA systems, it is not practical and realistic to apply NOMA jointly on all users in the network. 
Here, user pairing/grouping, in which the users of the network are split into many pairs/groups and NOMA is implemented within each pair/group, is an interesting research alternative \cite{ding2016impact}. 
This, from the perspective of securing untrusted users in a NOMA system \cite{basepaper}, opens up a novel research  challenging of finding which users of the system should be grouped together?

\subsection{State-of-the-Art}
To derive maximum benefits from NOMA, identifying the best user pairing strategy has been an active research direction. In \cite{ding2016impact}, user pairing for two scenarios, i.e., NOMA with fixed power allocation and cognitive-radio inspired NOMA was studied. In \cite{8016604},  user pairing in cognitive-radio based NOMA system was investigated, where the conventional distributed matching algorithm was applied. In \cite{user_pairing}, two user pairing strategies were proposed to maximize the ergodic sum capacity of the system. In \cite{8269086}, a novel user pairing strategy was provided to improve the performance of the weak user in the presence of two strong users. A virtual user pairing NOMA strategy was proposed in \cite{7600382} to utilize spectrum efficiently.

Utilizing the concept of PLS, various works have investigated the secrecy performance of NOMA systems in the presence of untrusted users. An untrusted users' network is a hostile but realistic situation where users do not have mutual trust. Therefore, they focus on securing their own data from other users, which leads to complex resource allocations \cite{9188014}, \cite{9324786}. In \cite{basepaper}, strong and weak users of a two-user NOMA system were assumed to be trusted and untrusted, respectively, and the secrecy outage probability was analyzed for the strong user against the weak user. In \cite{7833022}, assuming strong and weak users of a multiple-input single-output NOMA system as trusted and untrusted, respectively, the sum secrecy rate for the strong user was analyzed. In \cite{8945391}, two optimal relay selection schemes were proposed to attain secure communication for a strong user in the presence of an untrusted weak user. 

\subsection{Motivation and Contributions}
We observed that \cite{ding2016impact}, \cite{8016604}-\cite{7600382}  worked on the user pairing problem in NOMA. Whereas, \cite{basepaper}, \cite{7833022}, \cite{8945391} analyzed PLS performance of an untrusted NOMA system. Thus, PLS and user pairing for NOMA, have been studied independently in the literature. However, to our best knowledge, exploring the best user pairing scheme from the perspective of improving the secrecy rate in an untrusted NOMA system has not been considered yet. 
Addressing this, our key contributions are:
\begin{itemize}
\item Assuming a downlink communication system, we focus on finding out the best user pairing scheme from the perspective of improving the secrecy rate of the stronger user while fulfilling the QoS demands of the weaker user. 
\item  Firstly, we consider a two-user NOMA pair and find out a closed-form solution of optimal power allocation that can maximize the secrecy rate for the strong user while ensuring a minimum data rate guarantee to the weak user. 
\item Then, we consider a four-user system and explore the best pairing scheme for three different scenarios. The first one maximizes the secrecy rate of the strong user, the second scenario maximizes the weak uesr's data rate, and the third one maximizes the secrecy rate of the strong user while satisfying weak user's minimum rate requirement. 
\item Lastly, numerical results  verify the correctness of the analytical derivations and present valuable insights on the secrecy rate performance achieved through our schemes.
\end{itemize}
 
\section{Secure NOMA  with Untrusted Users}

\subsection{System Model and NOMA Transmission Principle} \label{sec_system_model}
We consider the downlink system with one base station and $N$ users. The $i$-th user of the system is denoted as $U_{i}$, where $i\in\mathcal{N}=\{1,2,...,N\}$. Each node in the system has a single antenna. The channel coefficient between the base station and $U_{i}$, which is assumed to be Rayleigh distributed, is denoted by $h_{i}$. 
The corresponding channel power gain $|h_i|^2$ between base station and $U_{i}$ experiences exponential distribution with mean $\lambda_{i}= L_{p} d_{i}^{-e}$, where $L_{p}$, $e$, and $d_{i}$, respectively, denote path loss constant, path loss exponent and distance between BS and $U_i$. Without loss of generality, the channel power gains are assumed to be sorted as $|h_{1}|^{2} > |h_{2}|^{2} > . . . >|h_{N}|^{2}$.

To achieve high spectral efficiency, users are multiplexed in the same resource block of a NOMA system. Considering an interference-limited NOMA system \cite{ding2016impact}, we restrict to the simplest case of two users in a pair. 
In general, the user with a better channel gain is paired with the user having poorer one. 
However, the key question is - which two users should be selected to be paired together? For the sake of our discussion, we consider two paired users as $U_{m}$ and $U_{n}$, out of all $N$ users with $|h_m|^2 >|h_n|^2$, where $m,n \in \mathcal{N}, n>m$.  Using power domain NOMA, for pair of users $U_{m}$ and $U_{n}$, the base station transmits a superimposed signal  $x=\sqrt {\alpha _{m}P_t}x_{m}+\sqrt {\alpha _{n}P_{t}}x_{n}$, where $x_{i} \forall i \in \{m,n\}$ is the signal associated with the $i$-th user,  $P_{t}$ is base station's transmission power, $\alpha_{i}$ denotes the coefficient of signal power for $U_{i}$, satisfying  $\alpha_{m} + \alpha_{n} = 1$.  We assume received noise to be additive white Gaussian with mean 0 and variance $\sigma^{2}$  for both users.  

\subsection{Achievable Secrecy Rates at Users}
In NOMA, the stronger user $U_{m}$ first decodes and removes $x_n$ by exploiting successive interference cancellation, and then decodes $x_m$. We assume that the weaker user is an untrusted user and may try to get the message of the stronger user. Thus, the weaker user $U_{n}$ will first decode its own signal $x_n$, and then may decode $x_m$ \cite{basepaper}. Using superscript $(m,n)$ to denote the indices of the paired users $U_{m}$ and $U_{n}$, where $m,n \in \mathcal{N}, n>m$, the received signal-to-interference-plus-noise-ratio (SINR) at $U_a$, when signal of $U_a$ is decoded by $U_b$, where $(a=m,n)$ and $(b=m,n)$, is denoted as $\Gamma_{ab}^{(m,n)}$. As a result, for the pair $(m,n)$, the SINRs can be obtained as  
\begin{align}\label{SINRT}
&\Gamma^{(m,n)}_{nm} = \frac{\alpha_{n}|h_m|^{2}}{\alpha_{m}|h_m|^{2} + \frac{1}{\rho_{t}}}, \quad 
&\Gamma^{(m,n)}_{nn} = \frac{ \alpha_{n}|h_n|^{2}}{\alpha_{m}|h_n|^{2} + \frac{1}{\rho_{t}}},\nonumber\\
&\Gamma^{(m,n)}_{mm} = \alpha_{m}|h_m|^{2}\rho_{t},\quad
&\Gamma^{(m,n)}_{mn} = \alpha_{m}|h_n|^{2}\rho_{t}.  
\end{align} 
Here, $\rho_{t} = \frac{P_{t}}{ \sigma^{2}}$ indicates the base station transmit signal-to-noise ratio (SNR). Further, to analyze the secrecy performance from the perspective of PLS, the achievable secrecy rate for $U_{a}$, where $(a=m,n)$ can be mathematically expressed as 
\begin{align}\label{secrecy_rate}
R_{sa}^{(m,n)} = R_{aa}^{(m,n)} - R_{ab}^{(m,n)}, a \neq b.
\end{align}
In \eqref{secrecy_rate}, $R_{ab}^{(m,n)}$ can be given by Shannon's capacity formula as 
\begin{align}\label{info_rate}
R_{ab}^{(m,n)} = \log_{2}(1 + \Gamma^{(m,n)}_{ab}).
\end{align}

\subsection{Feasibility Conditions for Positive Secrecy Rate}
The key idea of achieving a positive secrecy rate for a user is to ensure that the data rate over the desired link is higher than that of the eavesdropper's link. Mathematically, for $R_{sa}^{(m,n)}>0$,  $R_{aa}^{(m,n)}>R_{ab}^{(m,n)}$, simplified as $\Gamma^{(m,n)}_{aa}>\Gamma^{(m,n)}_{ab}$, must be satisfied. Lemma \ref{Lemma1} provides a key result on achievable secrecy performance in an untrusted NOMA system.
\begin{Lemma}\label{Lemma1}
For every pair of users in an untrusted NOMA system, the message of the stronger user is always safe from the untrusted weaker user. In comparison, the data of the weaker user is not secured from the stronger user.
\end{Lemma}
\begin{IEEEproof}
For users $U_{m}$ and $U_{n}$, we assume $|h_m|^2 >|h_n|^2$, where $m,n \in \mathcal{N}, n>m$. To check positive secrecy rate for stronger user $U_m$ against $U_{n}$, using \eqref{SINRT}, we solve $\Gamma^{(m,n)}_{mm}>\Gamma^{(m,n)}_{mn}$, and get a feasible condition $|h_{m}|^{2}>|h_{n}|^{2}$. This indicates that always positive secrecy rate is achieved for the strong user $U_{m}$ against the weak user $U_{n}$. Similarly, to achieve positive secrecy rate for $U_{n}$, we solve $\Gamma^{(m,n)}_{nn}>\Gamma^{(m,n)}_{nm}$ and obtain $|h_{n}|^{2}>|h_{m}|^{2}$, using \eqref{SINRT}, which is an infeasible condition. Thus, data of $U_{n}$ cannot be secured from  $U_{m}$.
\end{IEEEproof}

\section{Proposed Optimal Power Allocation Policy}\label{sec:PA}

To maximize the secrecy rate of the stronger user $U_{m}$, under the constraint of fulfilling QoS demands of the weaker user $U_{n}$, the optimization problem for user pair $(m,n)$ can be formulated using \eqref{secrecy_rate} and \eqref{info_rate}, as
\begin{align}
\mathcal{P}1:&\mathop {\mathrm {maximize}}\limits _{\alpha _{m}, \alpha_{n}} \quad R_{sm}^{(m,n)}, \quad\text{subject to}\nonumber \\ 
  &\mathcal{C}1\!: R_{nn}^{(m,n)}\geq R_{\mathrm{th}}, \;\mathcal{C}2\!: \alpha _{m}, \alpha_{n} > 0, \;
  \mathcal{C}3\!: \alpha_{m} + \alpha_{n} = 1, \nonumber
\end{align}
where $R_\mathrm{th}$ denotes the targeted data rate to be guaranteed for the weaker user $U_{n}$.
The problem $\mathcal{P}$1 can be reformulated via considering the power allocation constraints $\mathcal{C}2$ and $\mathcal{C}3$, which can be simplified in terms of $\alpha_{m}$ and $\alpha_{n}=1-\alpha_{m}$. Thus, the reformulated optimization problem can be stated as
\begin{align}
\mathcal{P}2: \mathop {\mathrm {maximize}}\limits _{\alpha _{m}} \; &R_{sm}^{(m,n)}, \quad\;\text{subject to}\quad  \mathcal{C}1, \mathcal{C}4: 0<\alpha _{m} < 1. \nonumber 
\end{align}

The optimal solution of the problem $\mathcal{P}2$ is given below.

\begin{Lemma}\label{Lemma2}
The optimal $\alpha_{m}$ that maximize the secrecy rate $R^{(m,n)}_{sm}$ of $U_{m}$ against $U_{n}$, under the minimum data rate (QoS) guarantee  for $U_{n}$ and definition $\Pi\stackrel{\Delta}{=}2^{R_{\mathrm{th}}}$, can be written as
\begin{align}\label{optimal_alpha}
\alpha_{m}^{*} = \frac{|h_n|^2 \rho_{t}- \Pi+1}{\Pi|h_n|^2 \rho_{t}}. 
\end{align} 
\end{Lemma}

\begin{IEEEproof}\label{proof_Lemma_2}
From, $\eqref{SINRT}$, $\eqref{secrecy_rate}$ and $\eqref{info_rate}$, we see that secrecy rate $R^{(m,n)}_{sm}$ is a function of $\alpha_{m}$. We check the first order derivative of $R^{(m,n)}_{sm}$ with respect to $\alpha_{m}$, which is given by  
\begin{align}\label{derivative}
\frac{\text{d}R^{(m,n)}_{sm}}{\text{d}\alpha_{m}}=\dfrac{\left(|h_m|^2-|h_n|^2\right)\rho_{t}}{\ln\left(2\right)\left(1+\alpha_{m}|h_m|^2\rho_{t} \right)\left(1+\alpha_{m}|h_n|^2\rho_{t}\right)}.
\end{align}
It can be easily observed from \eqref{derivative}, that $\frac{\text{d}R^{(m,n)}_{sm}}{\text{d}\alpha_{m}}>0$, since $|h_m|^2>|h_n|^2$ is assumed in Section \ref{sec_system_model}. Thus, $R^{(m,n)}_{sm}$ is an increasing function with respect to $\alpha_{m}$. Thus, to obtain maximum secrecy rate for $U_{m}$, $\alpha_{m}$ should be maximum.    

On the other hand, the range of the $\alpha_{m}$ can be obtained by solving the constraint $\mathcal{C}1$ directly. The constraint $\mathcal{C}1$ which is given as $R^{(m,n)}_{nn} \ge R_{\mathrm{th}}$ can be simplified using \eqref{info_rate} as $\Gamma^{(m,n)}_{nn} \ge (\Pi-1)$. Simplifying it further, boundary condition on $\alpha_{m}$ is
\begin{align}\label{boundary_constraint}
\alpha_{m} \le \frac{|h_n|^2 \rho_{t}- \Pi+1}{\Pi|h_n|^2 \rho_{t} } \triangleq{\alpha^u}.
\end{align}

Thus, the optimal solution $\alpha_{m}^*$ of $\mathcal{P}2$ is  $\alpha_{m}^* = \alpha^u$.
\end{IEEEproof}

Since, $\alpha_{m}^{*}$ is the power allocation coefficient, we investigate the feasibility of optimal solution $\alpha_{m}^{*}$ through Proposition \ref{Proposition1}.

\begin{Proposition}\label{Proposition1}
With $\alpha_{m}^{*} \in (0, 1)$  being a real number, the minimum rate guarantee  $R_{\mathrm{th}}$ of the weaker user is bounded as: $0<R_{\mathrm{th}}<\log_2(|h_{n}|^{2}\rho_{t}+1)$. 
\end{Proposition}
\begin{IEEEproof}
Solving $\alpha_{m}^{*}>0$, we obtain a feasibility condition $\Pi<|h_{n}|^{2}\rho_{t}+1$, which can be further simplified as $R_{\mathrm{th}}<\log_2(|h_{n}|^{2}\rho_{t}+1)$, since $\Pi=2^{R_{\mathrm{th}}}$. Similarly, on solving $\alpha_{m}^{*}<1$, we obtain $\Pi>1$, which gets simplified to $R_{\mathrm{th}}>0$.  Thus, for $\alpha_{m}^{*}$ to be in between 0 and 1, the condition $1<\Pi<|h_{n}|^{2}\rho_{t}+1$ must be fulfilled. Simplifying the conditions in terms of $R_{\mathrm{th}}$, we obtain  $0<R_{\mathrm{th}}<\log_2(|h_{n}|^{2}\rho_{t}+1)$. 
\end{IEEEproof}

\section{User-Pair Selection} \label{Section4}
To analyze the impact of different user-pairs on the secrecy performance achieved in an untrusted NOMA system, we consider a network with four users and choose a pair to apply NOMA while keeping other two users unpaired. 
We start with discussing the possible pairs to eventually find best user pair.

\subsection{Possible User-Pairs in a 4-User System}
With the intention of choosing two users out of four for making a pair, there are $^4C_2 = 6$ ways to form user pairs. Let us define all possible pairs by a set $\mathbb{S}=\{(m,n)| m,n\in \mathcal{N},  n>m\}$. Here, $(m,n)$ denotes the indices of the paired users $U_{m}$ and $U_{n}$. Thus, the set of all possible pairs for a system with four users can be written as $\mathbb{S}=\{(1,2), (1,3), (1,4), (2,3), (2,4), (3,4)\}$. 

\begin{table}
\centering
\caption{Order of pairs in terms of secrecy achieved at strong user ($\mathcal{A}$ and $\mathcal{B}$) and data rate achieved at the weak user ($\mathcal{C})$}\scriptsize\vspace{-3mm}
\begin{tabular}{|l|l|l|l|}
\hline
 & $\mathcal{A}$ & $\mathcal{B}$ & $\mathcal{C}$ \\\hline
Pair &  $|h_{i}|^{2} \!=\! (N\!-\!i+1)|h_{N}|^{2}$  & $|h_{i}|^{2}\! =\! \frac{|h_{1}|^{2}}{i}$ & $|h_{i}|^{2}>|h_{i+1}|^{2} $\\\hline
(1,2) & VI (Worst) & III & I (Best)\\\hline
(1,3) & III & II & II\\\hline
(1,4) & I (Best) & I (Best) & III\\\hline
(2,3) & V & V & II\\\hline
(2,4) & II & IV & III\\\hline
(3,4) & IV & VI (Worst) & III\\\hline
\end{tabular}
\label{tab:label}
\end{table}

\subsection{Best User-Pair Selection}
To find the best pair out of set $\mathbb{S}$, we compare each pair with the other pair one by one.  For example, pair $(m,n)\in \mathbb{S}$ is compared with the pair $(k,l)\in \mathbb{S}$, where $(m,n)\neq(k,l)$. Considering all possible combinations, we observe that the total possible comparisons are $^6C_2 = 15$. In the following, we investigate the best pair under three different scenarios. 

\subsubsection{Best User-Pair to Maximize Secrecy Rate of the Strong User}\label{Section4A}
In this case, we focus on maximizing the secrecy rate of the strong user only without focusing on the data rate demands of the weak user. Therefore, we neglect the QoS constraint of the weak user. We note from \eqref{derivative} that the secrecy rate $R_{sm}^{(m,n)}$ for $U_{m}$ is an increasing function of $\alpha_{m}$. It means that, to achieve maximum secrecy rate for $U_{m}$,  $\alpha_{m}$ should be maximum. Therefore, in this case, we take $\alpha_{m}=1$ and $\alpha_{n}=0$. Below a key result on the best pair from a secrecy perspective is provided through Proposition \ref{Proposition_2}. 
\begin{Proposition}\label{Proposition_2}
Ignoring the data rate demands of the weak user, the optimal secrecy rate performance for the strong user can be achieved by selecting those two users for which the channel power gains difference is maximum.
\end{Proposition}
\begin{IEEEproof}
Considering $\alpha_{m}=1$ for the pair $(m,n)\in \mathbb{S}$, the secrecy rate, using \eqref{SINRT}, and \eqref{secrecy_rate},  can be expressed as $R_{sm}^{(m,n)} = \log_2\Big(\frac{1+|h_{m}|^{2}\rho_t}{1+|h_{n}|^{2}\rho_t}\Big)$. It can be noted that $R_{sm}^{(m,n)}$ increases with $|h_{m}|^{2}$, and decreases with $|h_{n}|^{2}$. Thus, the secrecy rate of the strong user is dependent on the ratio of channel power gains of the strong user and the weak user.  Therefore, when $|h_{1}|^{2} > |h_{2}|^{2} >|h_{3}|^{2}  >|h_{4}|^{2}$, the user with the best channel condition, i.e., $U_{1}$, should be paired with the user having the worst channel condition, i.e., $U_{4}$. 
\end{IEEEproof}

Next, we focus on obtaining the ordered sequence of user pairs in terms of decreasing the secrecy rate performance of the strong user from the highest to the lowest. To investigate this order, we consider a specific scenario, where channel gains are assumed to follow $|h_{i}|^{2} = (N-i+1)|h_{N}|^{2}$, $\forall i\in\mathcal{N}$. In this regard, a key result is provided in Proposition \ref{Proposition_3}.

\begin{Proposition}\label{Proposition_3}
For an untrusted NOMA system, where channel power gains are such that $|h_{i}|^{2} = (N-i+1)|h_{N}|^{2}$, $\forall i\in\mathcal{N}$, the highest to the lowest order of the secrecy rate achieved for the strong user in all possible pairs is given as 
\begin{equation}
R_{s1}^{(1,4)}>R_{s2}^{(2,4)}>R_{s1}^{(1,3)}>R_{s3}^{(3,4)}>R_{s2}^{(2,3)}>R_{s1}^{(1,2)}
\end{equation}
\end{Proposition}
\begin{IEEEproof}
The secrecy rate of a user in each pair is in the form of $\log_2\big(\frac{1+A}{1+B}\big)$. On comparing two pairs, we notice that a given secrecy rate is higher than its counterpart if either $A$ is higher and/or $B$ is lower, with other respective parameters being the same. We investigate each comparison one by one and find out the feasibility conditions. Like, we compare two user pairs $(1,4)$ and $(2,4)$. The secrecy rate for $U_{1}$ in the case of $(1,4)$ with $\alpha_{1}=1$ and $|h_1|^{2}=4|h_4|^{2}$, can be given as
\begin{equation}\label{1,4}
R_{s1}^{(1,4)} \!=\! \log_2\Bigg(\frac{1+|h_1|^{2}\rho_{t}}{1+|h_4|^{2}\rho_{t}}\Bigg) = \log_2\Bigg(\frac{1+4|h_4|^{2}\rho_{t}}{1+|h_4|^{2}\rho_{t}}\Bigg).
\end{equation}
Similarly, the secrecy rate for $U_{2}$ in the pair $(2,4)$, with $\alpha_{2}=1$ and $|h_2|^{2}=3|h_4|^{2}$, can be given as
\begin{equation}
R_{s2}^{(2,4)} \!=\! \log_2\Bigg(\frac{1+|h_2|^{2}\rho_{t}}{1+|h_4|^{2}\rho_{t}}\Bigg) = \log_2\Bigg(\frac{1+3|h_4|^{2}\rho_{t}}{1+|h_4|^{2}\rho_{t}}\Bigg).
\end{equation}
Here, $A=4|h_4|^{2}$ for the user-pair $(1,4)$, while $A=3|h_4|^{2}$ in the user-pair $(2,4)$. On comparing, we find that $A$ is higher in user-pair $(1,4)$ than the $A$ in the user-pair $(2,4)$. Thus, it is clear that $R_{s1}^{(1,4)}>R_{s2}^{(2,4)}$. Similarly, we compare all other possible combinations and obtain the order of secrecy rate achieved at strong users in the user-pairs as $R_{s1}^{(1,4)}>R_{s2}^{(2,4)}>R_{s1}^{(1,3)}>R_{s3}^{(3,4)}>R_{s2}^{(2,3)}>R_{s1}^{(1,2)}$.
\end{IEEEproof}

\begin{Remark}
Since the highest to the lowest order of the secrecy rate performance obtained at the strong user is dependent on the relative channel power gains of the users (refer Proposition \ref{Proposition_3}), the order gets changed if the power gains get changed. For example, if the power gains are such that $|h_{i}|^{2} = \frac{|h_{1}|^{2}}{i}$, where $\forall i \in \mathcal{N}$, then the obtained order of the pairs in terms of secrecy rate is provided in Table \ref{tab:label}. 
\end{Remark}

 
\subsubsection{Best User-Pair to Maximize Data Rate of the Weaker User}\label{Section4B}
In this case, we intend to select the best user-pair to maximize the data rate of the weak user without worrying about the secrecy performance of the strong user. The first order derivative of $R_{nn}^{(m,n)}$ with respect to $\alpha_{m}$ is given as 
\begin{align}
\frac{\text{d}R^{(m,n)}_{nn}}{\text{d}\alpha_{m}}=\frac{-|h_{n}|^{2}(|h_{n}|^{2}\rho_{t}+1)\rho_{t}}{(\alpha_{m}|h_{n}|^{2}\rho_{t} + 1)^{2}},
\end{align}
which is negative. Thus, $R^{(m,n)}_{nn}$ is a decreasing function of $\alpha_{m}$. Therefore, in this case, we take the lower bound on $\alpha_{m}$, i.e., $\alpha_{m}^{(m,n)}=0$ and thus, $\alpha_{n}^{(m,n)}=1$. The best user-pair is presented through Proposition \ref{proposition_4} in the following.  
\begin{Proposition}\label{proposition_4}
Regardless of the secrecy of the strong user, the maximum data rate at the weak user is obtained when the strongest user $U_{1}$ is paired with the second strongest user $U_{2}$.
\end{Proposition}
\begin{IEEEproof}
Using \eqref{SINRT}, \eqref{info_rate}, $\alpha_{m}=0$ and $\alpha_{n}=1$, the data rate $R^{(m,n)}_{nn}$  of the weak user $U_{n}$ in a pair $(m,n)$, can be given as
\begin{align}\label{mn_datarate}
R_{nn}^{(m,n)} = \log_{2}(1 + \Gamma^{(m,n)}_{nn})=\log_{2}(1 + |h_{n}|^{2}\rho_{t}).
\end{align}
$R_{nn}^{(m,n)}$ is an increasing function of $|h_{n}|^{2}$. Thus, while comparing two pairs, the pair having better channel condition of the weak user will provide a better data rate for the weak user. Since, we assume a system where $|h_{2}|^{2} >|h_{3}|^{2}  >|h_{4}|^{2}$, it is clear that $(1,2)$ will be the optimal user-pair. 
\end{IEEEproof}

\begin{Remark}
Based on Proposition \ref{proposition_4}, the highest to the lowest order of the user-pairs in terms of the data rate performance of the weak user has been presented in Table \ref{tab:label}. Since the ordering is based on the rate of the weak user, the ordering does not change with a change of the strong user.  
\end{Remark}


\subsubsection{Best User-Pair for QoS-Aware Secure NOMA}
Lastly, we consider the third case, where the goal is to find out the best user pair that maximizes the secrecy rate of the strong user while ensuring QoS demands of the weak user. To fulfil the minimum threshold rate requirements of the paired weak user in each user-pair, we consider the optimal power allocation as obtained in Section \ref{sec:PA}. Here, we consider a specific scenario, where channel power gains are assumed to follow $|h_{i}|^{2} = (N-i+1)|h_{N}|^{2}$, $\forall i\in\mathcal{N}$. To investigate the order of the pairs, we compare each pair with another, one by one. For example, let us compare two pairs $(2,4)$ and $(1,3)$. In pair $(2,4)$ secrecy rate for $U_{2}$ with $|h_{2}|^{2}=3|h_{4}|^{2}$, is  given as
\begin{eqnarray}\label{2,4}
R_{s2}^{(2,4)}\!\!=\! \log_2\left(\frac{1+\alpha_{2}^{*}|h_2|^{2}\rho_{t}}{1+\alpha_{2}^{*}|h_4|^{2}\rho_{t}}\right) \!=\! \log_2\left(\frac{1+\alpha_{2}^{*}3|h_4|^{2}\rho_{t}}{1+\alpha_{2}^{*}|h_4|^{2}\rho_{t}}\right)\!
\end{eqnarray}
where $\alpha_{2}^{*}=\frac{|h_4|^2 \rho_{t}- \Pi+1}{\Pi|h_4|^2 \rho_{t}}$ is the optimal power allocation from \eqref{optimal_alpha}.
Similarly, the secrecy rate for $U_{1}$ in the pair $(1,3)$, with $|h_1|^{2}=4|h_4|^{2}$ and $|h_3|^{2}=2|h_4|^{2}$, can be given as $R_{s1}^{(1,3)}= \log_2\Big(\frac{1+\alpha_{1}^{*}4|h_4|^{2}\rho_{t}}{1+\alpha_{1}^{*}2|h_4|^{2}\rho_{t}}\Big)$, 
with $\alpha_{1}^{*}=\frac{|h_3|^2 \rho_{t}- \Pi+1}{\Pi|h_3|^2 \rho_{t}}=\frac{2|h_4|^2 \rho_{t}- \Pi+1}{\Pi|2h_4|^2 \rho_{t}}$. To find out, which pair performs better in terms of secrecy rate of the strong user, we solve $ R_{s2}^{(2,4)}>R_{s1}^{(1,3)}$, resulting in a condition: $R_{\mathrm{th}}<\log_{2}\big(1+\frac{2|h_{4}|^{2}|h_{4}|^{2}\rho_{t}^{2}}{1+3|h_{4}|^{2}\rho_{t}}\big)$.

It shows that $(2,4)$ provides more secrecy rate than $(1,3)$ with a suitable constraint on threshold rate $R_{\mathrm{th}}$. In a similar manner, bounds on threshold rate can be obtained for other comparisons as well. However, in some of the comparisons, we also get a feasible condition that indicates that a pair is always better than the other pair. For example, to compare two  user-pairs $(1,4)$ and $(3,4)$, we solve $R_{s1}^{(1,4)}>R_{s3}^{(3,4)}$, and get a feasible condition $|h_{4}|^{2}\rho_{t}+1$. It means that pair $(1,4)$ always performs better in terms of secrecy rate for the strong user than the pair $(3,4)$ Thus, it can be concluded that in this case, the comparison between two pairs is dependent on the channel power gain conditions and threshold rate. Therefore, we observe the order of the pairs with optimal power allocation for different channel power gain ratios via simulations.

\section{Numerical Results}
Here, numerical results under various system settings are presented to validate the proposed results and present useful insights on the system performance. We consider a downlink system with four users, where the base station communicates with two users in a pair. Noise signal at both users is assumed to follow Gaussian distribution with an average noise power of -120 dBm. Besides, $L_{p}=1$, $e=3$, $R_{\mathrm{th}}=0.5$, and $P_{t}=1$ mW are considered.  Simulation results have been averaged over $10^6$ randomly generated channel realizations.


\begin{figure}[!t]
\centering
 \includegraphics[scale=.31]{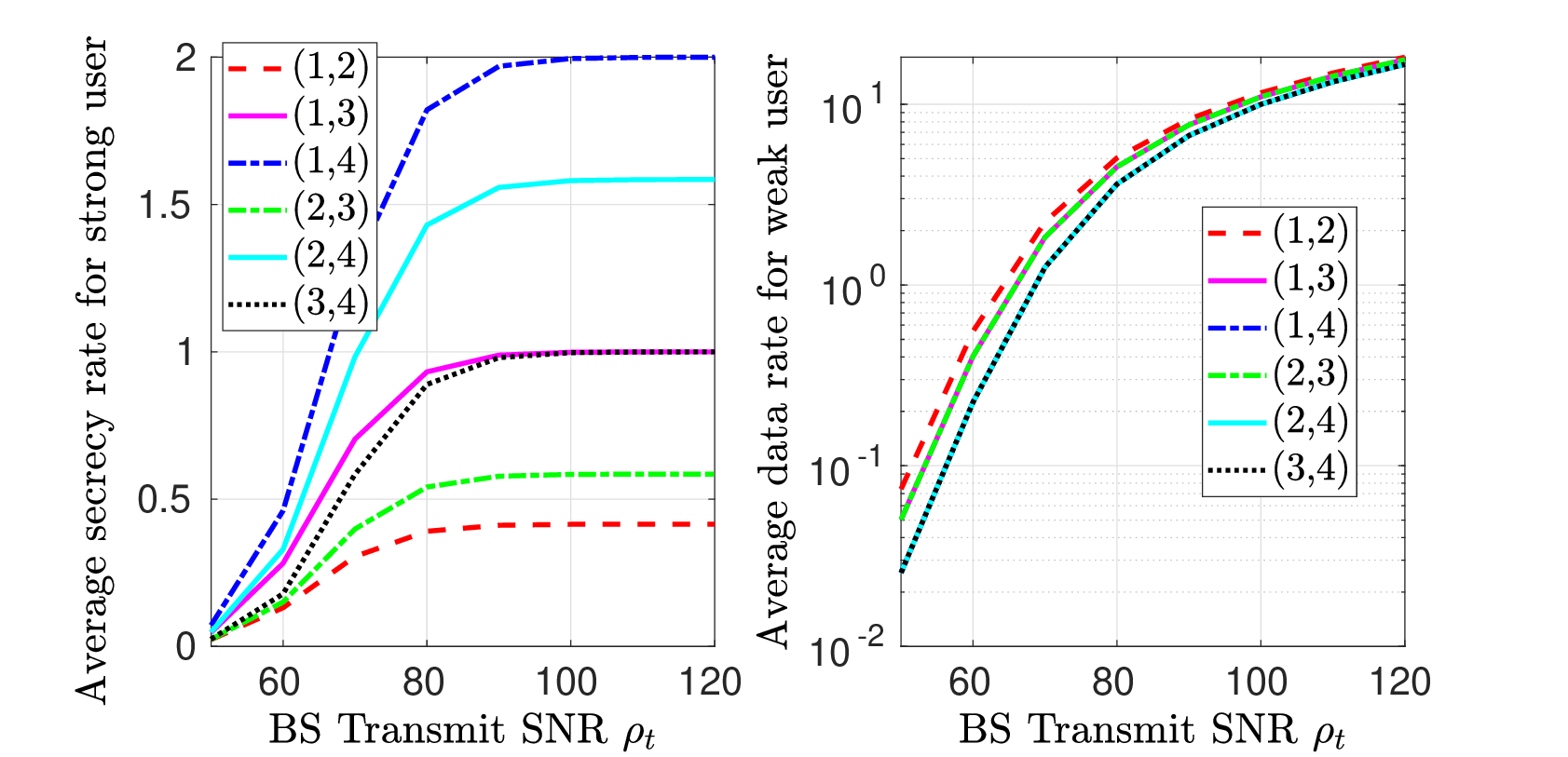}\vspace{-3.5mm}
\caption{Validation of order of user-pairs with the (a) average  secrecy rate of strong user with $\alpha_{m}=1$ and $\alpha_{n}=0$ and (b) average data rate for the weak user $\alpha_{m}=0$ and $\alpha_{n}=1$, for channels satisfying $|h_{i}|^{2} = (N-i+1)|h_{N}|^{2}$. }
\label{userpair_order}
\end{figure}

\begin{figure}[!t]
\centering
 \includegraphics[scale=.28]{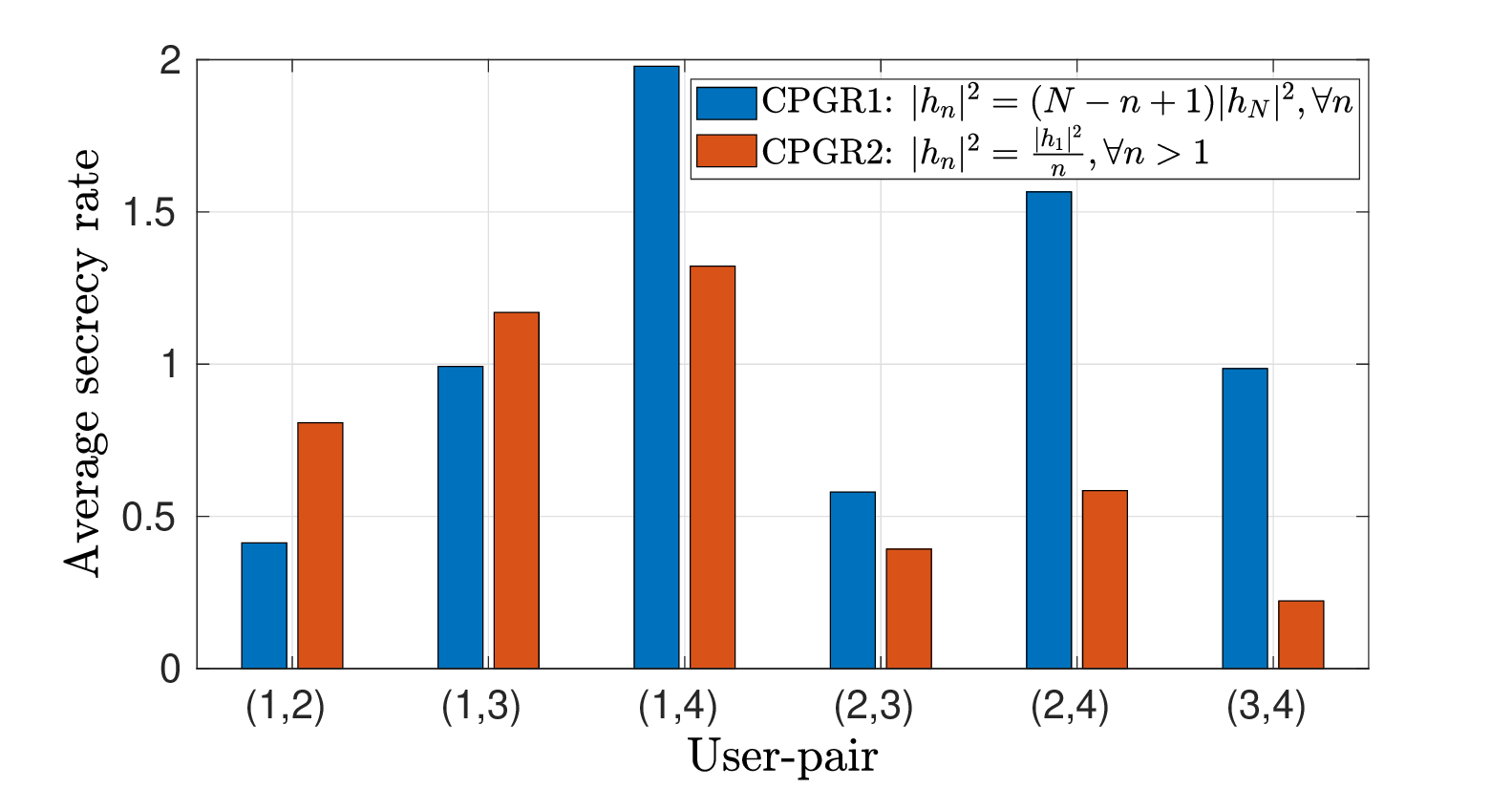}\vspace{-3.5mm}
\caption{Impact of different channel power gain ratios on average secrecy rate performance with optimal power allocation for various user-pairs. }
\label{channel_power_gain_impact}\vspace{-2mm}
\end{figure}

\begin{figure}[!t]
\centering
 \includegraphics[scale=.28]{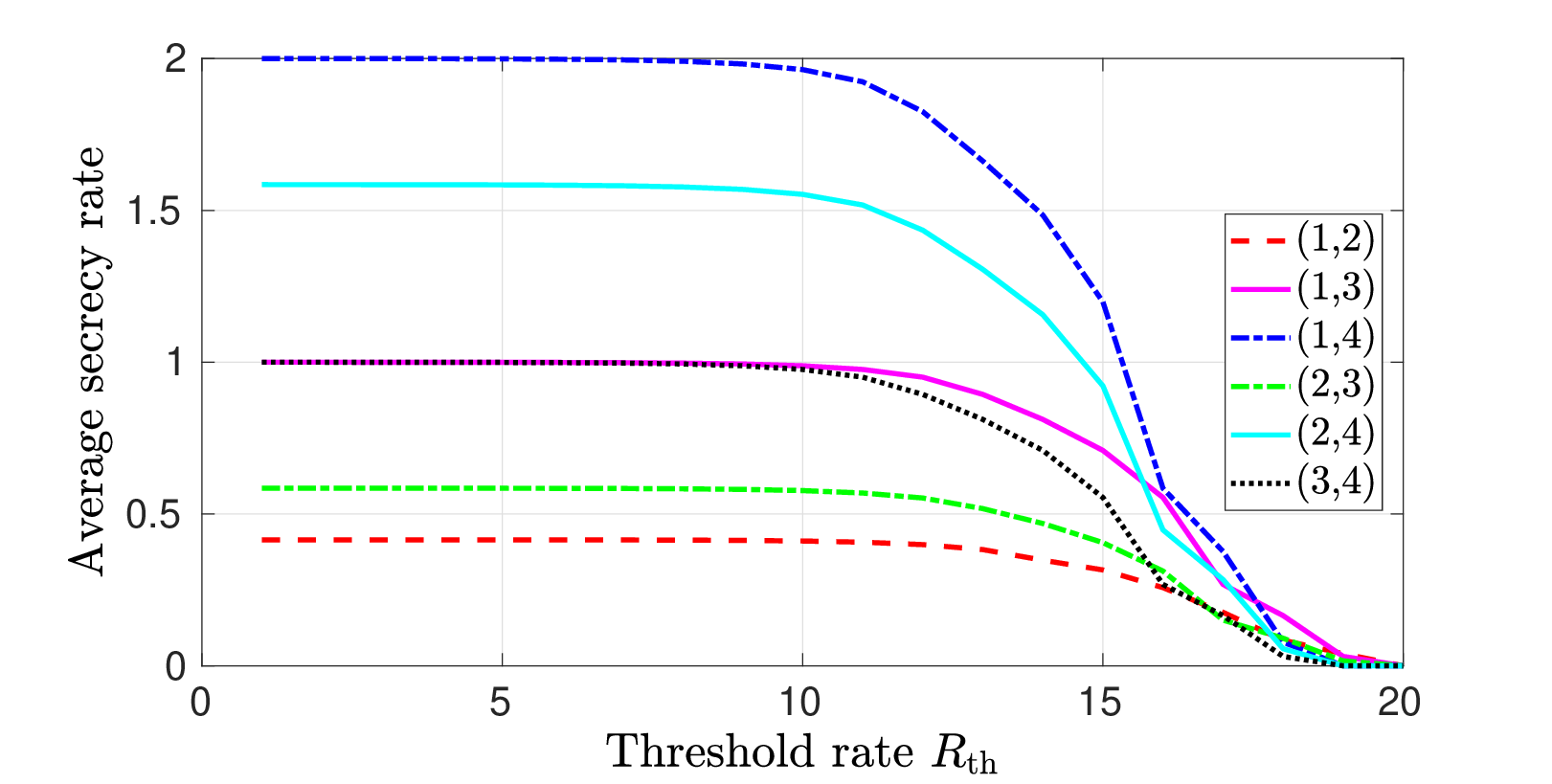}\vspace{-3.5mm}
\caption{Impact of different threshold rate requirements of a weak user on the average  secrecy rate of the strong user for  $|h_{i}|^{2} = (N-i+1)|h_{N}|^{2}$. }
\label{Threshold_rate}\vspace{-2mm}
\end{figure}

\begin{figure}[!t]
\centering
 \includegraphics[scale=.28]{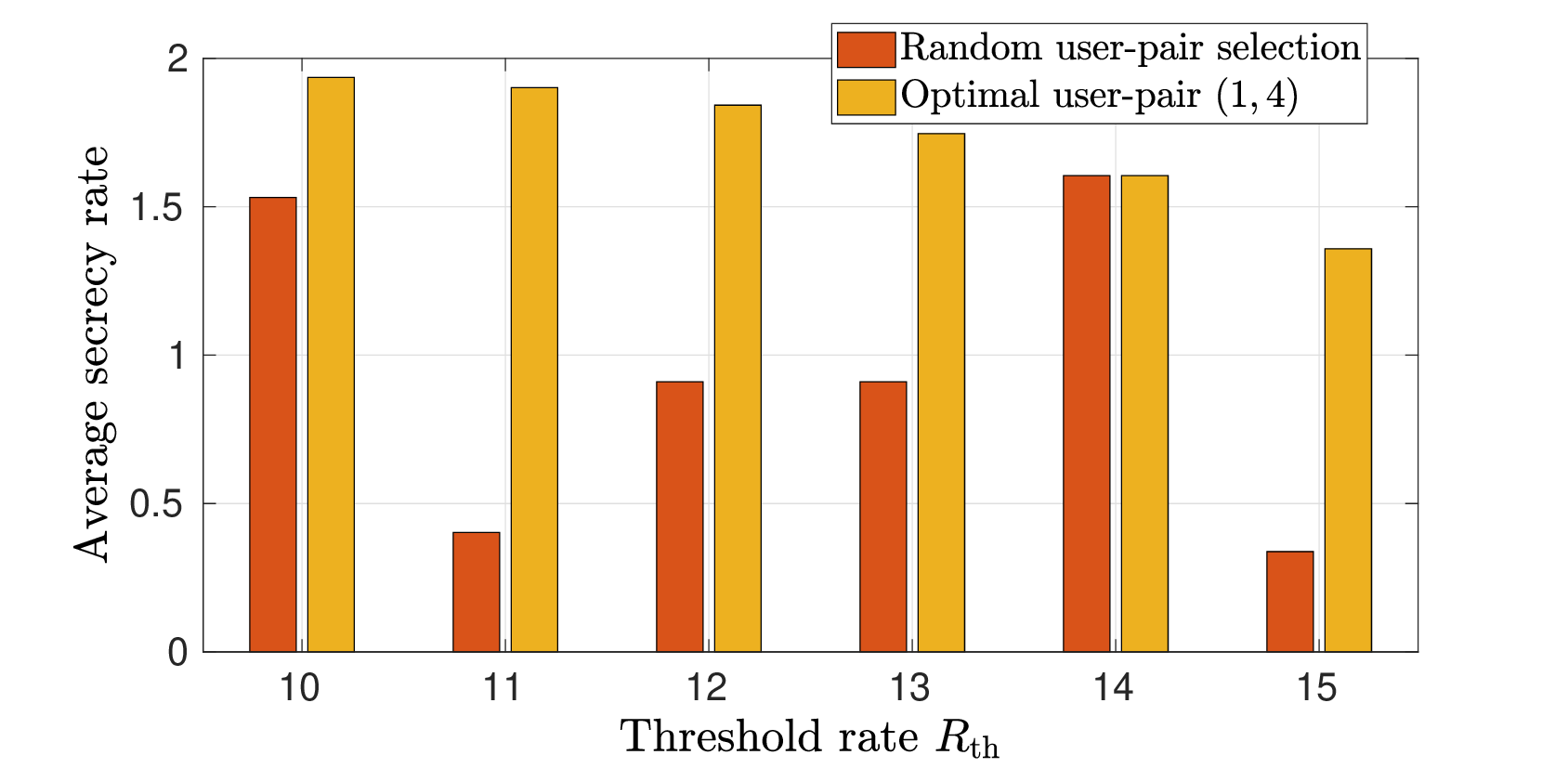}\vspace{-3.5mm}
\caption{Performance comparison of optimal user-pair with random user-pair selection for channel power gain conditions as  $|h_{i}|^{2} = (N-i+1)|h_{N}|^{2}$. }\vspace{-1.5mm}
\label{performance_comparison}
\end{figure}

\subsection{Insights on Optimal User-Pair Selection and Order}
We first observe the probability of occurrence of different user pairings for channel power conditions as $|h_{1}|^{2} > |h_{2}|^{2} >|h_{3}|^{2}  >|h_{4}|^{2}$ under three different cases. Case 1: maximizing secrecy rate for strong user only, Case 2: maximizing data rate of weak user only, and Case 3: maximizing secrecy rate for strong user while satisfying QoS requirements of weak user. Results validate that $(1,4)$ is the optimal pair to maximize the secrecy rate for the strong user as proposed in Proposition \ref{Proposition_2}. Similarly, simulations also confirm that  $(1,2)$ is the optimal pair for maximizing the data rate for the weak user  as proposed in Proposition \ref{proposition_4}. Through numerical observations, here we see that for the third case also, with optimal power allocation to users, pair $(1,4)$ wins. Thus, we conclude that the optimal user pair solution differs with the objective. 

Through the results shown in Fig. \ref{userpair_order}(a) and Fig. \ref{userpair_order}(b), we validate  the order proposed in Proposition \ref{Proposition_3} and Section \ref{Section4B}, respectively. Here, Fig. \ref{userpair_order}(a) is plotted for the case when the secrecy performance of the strong user is focused, and, Fig. \ref{userpair_order}(b) is plotted for the second case focusing on the data rate demands for the weak user. 

\subsection{Impact of Key Parameters and Performance Comparison}
Fig. \eqref{channel_power_gain_impact} is plotted to highlight the impact of various channel power gain conditions on the optimal secrecy rate performance of the strong user. Here, we observe that even with optimal power allocation, the highest to the lowest order of the secrecy rate performance for different pairs varies with the relative channel power gain conditions. Fig. \ref{Threshold_rate} is plotted to present the impact of increasing threshold rate requirements of the weak user on the secrecy rate performance achieved by the strong user. It is clearly observed that on increasing the QoS demands, the secrecy rate performance of the strong user decreases. This happens because to fulfil the QoS demands of the weak user, more power needs to be allocated to the weak user, due to which the rate of decoding the strong user will also increase. As a result, the secrecy rate of the strong user decreases. 

To observe the performance improvement achieved by the selection of optimal user-pair for maximizing secrecy rate at strong user, Fig.  \ref{performance_comparison} demonstrates its performance comparison with random user-pair selection for various threshold rates. It can be easily observed that that optimal user-pair performs best over the random user-pair selection. Here, the average percentage improvement of about $112\%$ is obtained.

\section{Conclusions}
This work studied the impact of user-pair selection on improving the secrecy performance of an untrusted NOMA system. The best user pair to maximize the secrecy rate of the strong user while satisfying the QoS demands of the weaker user is identified. To solve this problem, first optimal power allocation in a two-user NOMA system is obtained, and then a generalized problem with four users is considered. Numerical results validated the correctness of the proposed results and presented key insights on the impact of various system parameters on the system secrecy performance.

\section*{Acknowledgement}
This work was supported in parts by the SERB, DST, under the Grant no. CRG!2021/002464, and TCS RSP Fellowship.
\bibliographystyle{IEEEtran}
\bibliography{ref}
\end{document}